\documentstyle[eqsecnum,twocolumn,aps,prd,epsfig]{revtex}
\tighten
\begin{document}

\def\beq{\begin{equation}}
\def\endeq{\end{equation}}
\def\bea{\begin{eqnarray}}
\def\endea{\end{eqnarray}}
\wideabs{

\title{Proof of Classical Versions of the Bousso Entropy Bound and of the
Generalized Second Law}

\author{\'Eanna \'E. Flanagan}
\address{
Newman Laboratory of Nuclear Studies, Cornell University,
Ithaca, NY 14853-5001.\\
}

\author{Donald Marolf}
\address{
Institute for Theoretical Physics, University of California, Santa
Barbara, CA 93106\\
Physics Department, Syracuse University, Syracuse, NY13244.}

\author{Robert M.~Wald}
\address{
Enrico Fermi Institute and Department of Physics, University
of Chicago\\ 
5640 S. Ellis Avenue, Chicago, Illinois 60637-1433.\\
}

\date{\today}

\maketitle
\begin{abstract}

Bousso has conjectured that in any spacetime satisfying Einstein's
equation and satisfying the dominant energy condition, the ``entropy
flux'' $S$ through any null hypersurface $L$ generated by geodesics
with non-positive expansion starting from some spacelike 2 surface
of area $A$ must satisfy $S \le A / 4 G \hbar$.  This conjecture reformulates
earlier conjectured entropy bounds of Bekenstein and also of Fischler
and Susskind, and can be interpreted as a statement of the so-called
holographic principle.  
We show that Bousso's entropy bound can be derived from either of two  
sets of hypotheses.  
The first set of hypotheses is (i) associated with each null surface 
$L$ in spacetime there is an entropy flux 4-vector $s^a_L$ whose
integral over $L$ is the entropy flux through $L$, and 
(ii) along each null
geodesic generator of $L$, we have $|s^a_L k_a| \le \pi (\lambda_\infty -
\lambda) T_{ab} k^a k^b / \hbar$, where $T_{ab}$ is the
stress-energy tensor, $\lambda$ is an affine
parameter, $k^a = (d / d\lambda)^a$, and $\lambda_\infty$ is the
value of affine parameter at the endpoint of the geodesic.
The second (purely local) set of hypotheses is (i)
there exists an absolute entropy flux 4-vector $s^a$ such that
the entropy flux through any null surface $L$ is the integral of
$s^a$ over $L$, and (ii) this entropy flux 4-vector obeys the
pointwise inequalities $(s_a 
k^a)^2 \le T_{ab} k^a k^b / (16 \pi \hbar^2 G)$ and $|k^a k^b \nabla_a
s_b| \le \pi T_{ab} k^a  
k^b / (4 \hbar)$ for any null vector $k^a$.
Under the first set of hypotheses, we also show that a stronger entropy
bound can be derived, which directly implies the generalized second
law of thermodynamics.

\end{abstract}
}

\section{Introduction and Summary}
\label{s:intro}

\subsection{Background and Motivation}
\label{s:background}

In recent years, a number of independent universal entropy bounds have
been postulated to hold for arbitrary systems. The first such bound
was conjectured by Bekenstein, who proposed that the entropy $S$ and
energy $E$ of any matter put into a box must obey
\cite{Bek73}
\begin{equation}
S/E \leq 2 \pi R,
\label{S/E}
\end{equation}
where $R$ denotes some suitable measure of the size of the
box. [Throughout this paper, we use units with $G = c = \hbar = k =
1$.] The original motivation for the bound (\ref{S/E}) was the belief
that it is necessary for the validity of the generalized second law (GSL) of
thermodynamics, which states that in all physical processes the
generalized entropy
\begin{equation} 
S' = S + S_{\rm bh}
\end{equation}
must always increase, where $S$ is the entropy of matter outside of
black holes, $S_{\rm bh} = A_H/4$, and $A_H$ denotes the
total surface area of all black hole horizons.  Subsequently it was
shown \cite{uw82,uw83,pw99} that the bound
(\ref{S/E}) is not necessary for the validity of
the generalized second law\footnote{Very recently, Bekenstein \cite{buoy}
has used the
fact that the buoyancy formulas must be modified due to finite box size  
effects to again argue that a bound of the form (\ref{S/E}) is needed for
the validity of the GSL. However, we believe that an analysis of the type
given in \cite{uw83} could be used to show that no such entropy bound is
needed. Indeed, if a violation of the GSL could be obtained in any process
involving the quasi-static lowering of a box toward a black hole, then we
expect that it
should be possible to obtain a violation of the ordinary second law by a
similar quasi-static lowering of a box into a real star composed of
unconstrained thermal matter.}.  
In addition, the bound fails
when the number of species of particles is sufficiently
large\footnote{In the canonical ensemble, it is easy to show that the bound
(\ref{S/E}) also fails at sufficiently low temperatures for all systems
whose ground state energy vanishes. However, a detailed analysis of a
variety of systems given in Ref.\ \cite{review} provides strong evidence
that this failure does not occur in the microcanonical ensemble.  }.
Finally, it is far from clear what the precise meaning of
``$R$'' in the conjecture is supposed to be, particularly in curved
spacetime; in curved spacetime, it is also far from clear what ``E'' means. 
Nevertheless, a case can be made that the bound (\ref{S/E}) may
hold for all physically realistic systems found in nature; see
Ref.\ \cite{review} for further discussion.

More recently, an alternative entropy bound has been considered: the
entropy $S$ inside any region whose boundary has area $A$ must satisfy
\cite{others}
\begin{equation}
S \le A / 4.
\label{Sbound}
\end{equation} 
An argument given in Ref.\ \cite{Sus95} suggests that the bound
(\ref{Sbound}) should follow from the GSL together with the assumption
that the entropy of a black hole counts the number of possible
internal states of the black hole\footnote{The argument is attributed
to Bekenstein in Ref.\ \cite{Sus95} and 
goes as follows.  If black hole entropy counts the number    
of internal states of a black hole, then any system having $S \ge A/4$ is not 
a black hole.  Then, one would expect to be able to make that system
into a black hole with area $A$ by collapsing a sufficiently massive
spherical shell of matter around it.  In this process, it appears that
no entropy escapes, but this means that we convert an
$S \ge A/4$ system into a black hole of area $A$, violating the
generalized second law.  An antecedent to this argument can be found
in Ref.\ \cite{Bek74}.  For a counterargument, see Ref.\
\protect{\cite{Waldreview}}}.  In addition, 
when $E \alt R$, this bound would follow from the 
original Bekenstein bound (\ref{S/E}).  
The inequality (\ref{Sbound}), like
the bound (\ref{S/E}), can be violated if the number of massless
particle species is allowed to be arbitrarily large\footnote{
For example, consider $N$ free massless scalar fields in flat
spacetime, in a cube of edge length $L$ with Dirichlet boundary
conditions.  In the canonical ensemble, the thermal state with 
temperature $T$ with $T \ll 1/L$ has energy $E$ which scales as $E
\sim N \exp[-  \pi 
/ (L T)] / L$ and entropy $S$ which scales like $S \sim 
N \exp[- \pi / (L T)] / (L T)$.  For the system to be weakly
self-gravitating (a necessary condition for the flat-spacetime
analysis to be a good approximation) we must have $E = \varepsilon L$
for some $\varepsilon \ll 1$.  Using this restriction to solve for the
maximum allowed value of $N$ yields 
$
S/L^2 \sim \varepsilon / (L T),
$
which can be arbitrarily large. 
In the microcanonical ensemble, the violation of Eq.\ (\ref{Sbound})
for sufficiently  
large $N$ follows immediately from the fact that the density of states at
a fixed total energy $E$ grows unboundedly with $N$ at fixed $L$.
(Note, however, that Casimir energy has been ignored here.)}.  The
inequality (\ref{Sbound}) is  
related to the 
hypothesis known as the holographic principle, which states that the
physics in any spatial region can be fully described in terms of
degrees of freedom living on the boundary of that region, with one
degree of freedom per Planck area \cite{Tho93,Sus95}.  If the
holographic principle is correct, then since the entropy in any region
should be bounded above by the number of fundamental degrees of
freedom in that region, a bound of the form (\ref{Sbound}) should be
valid for all systems, including those with strong self-gravity.

As it stands, the bound (\ref{Sbound}) is ambiguous, since
the precise meaning of the ``bounding area'', $A$, has not been spelled
out.  In particular, note that any world tube can always be ``enclosed'' by a
two-surface of arbitrarily small area, since given any two-surface in
spacetime, there exists a two-surface of arbitrarily small area 
arbitrarily close to the original two-surface
(obtained by ``wiggling'' the original
two-surface suitably in spacetime).
However, very recently, a specific
conjecture of the form (\ref{Sbound}) was suggested by Bousso
\cite{Bou99b,Bou99c}, who improved an earlier suggestion of Fischler
and Susskind \cite{FisSus98,note33}.  Bousso showed that several example
spacetimes, including cosmological models and gravitational collapse
spacetimes, are consistent with his conjecture.

Bousso's conjecture is as follows.  Let $(M,g_{ab})$ be a
spacetime satisfying Einstein's equation and also the dominant energy
condition \cite{Wald}. Let $B$ be a connected 2 dimensional spacelike
surface in $M$.  Suppose that $k^a$ is a smooth null vector field on
$B$ which is everywhere orthogonal to $B$.  Then the expansion 
\beq
\theta = \nabla_a k^a 
\endeq 
of $k^a$ is well defined and is
independent of how $k^a$ is extended off $B$.  Suppose that $\theta
\le 0$ everywhere on $B$.  Let $L$ denote the null hypersurface
generated by the null geodesics starting at $B$ with initial tangent
$k^a$, where each null geodesic is terminated if and only if a caustic
is reached (where $\theta \to -\infty$), and otherwise is extended as
far as possible.  Then the entropy flux, 
$S_L$, through $L$ satisfies
\beq S_L \le A_B /4,
\label{Bousso1}
\endeq
where $A_B$ is the area of $B$.

There is a close relationship between Bousso's conjecture and the
generalized second law. 
Consider a foliation of the horizon of a black hole by spacelike
two-surfaces $B(\alpha)$, where $\alpha$ is a continuous label that
increases in the future direction (with respect to the time
orientation used to define the black hole).  
Let $A(\alpha)$ be the area of the
two surface $B(\alpha)$, and let $S(\alpha)$ be the total entropy that
crosses the horizon before the 2-surface $B(\alpha)$.  Then if one
assumes the ordinary second law, the GSL is equivalent 
to the statement that for any $\alpha_1 < \alpha_2$ we have
\beq
S(\alpha_2) - S(\alpha_1) \le {1 \over 4} \left[ A(\alpha_2) -
A(\alpha_1) \right].
\label{GSLeg}
\endeq
On the other hand, Bousso's entropy bound applied to the 2-surface
$B(\alpha)$---with $k^a$ taken to be the past directed normal to the
horizon, so that we have $\theta \leq 0$ on $B(\alpha)$ when the null
energy condition is satisfied---demands merely that
\beq
S(\alpha) \le {1 \over 4} A(\alpha) 
\label{Boussoeg}
\endeq
for all $\alpha$. Thus, Bousso's bound implies that the GSL holds for the
case when the 
initial time, $\alpha_1$, is taken to be the time
when the black hole is first formed [so that $S(\alpha_1) =
A(\alpha_1) = 0$]. In general, however, it is clear that the statement
(\ref{Boussoeg}) is weaker than the statement (\ref{GSLeg}).

This observation motivates a generalization of Bousso's conjecture.
Namely, if one allows the geodesics generating the hypersurface $L$ to
terminate at some spacelike 2-surface $B^\prime$ before coming to a
caustic or singularity, one can replace the conjectured inequality
(\ref{Bousso1}) by the condition
\beq
S_L \le {1 \over 4} \left[ A_B - A_{B^\prime} \right].
\label{Bousso2}
\endeq
It is clear from the above discussion that this
more general bound implies both the original Bousso entropy bound and
the GSL (assuming of course the validity of the ordinary second law).

In this paper we shall prove Bousso's entropy bound
(\ref{Bousso1}) under two independent sets of hypotheses concerning
the local entropy content of matter.  Furthermore, under the first set
of hypotheses, we will prove the more general entropy bound (\ref{Bousso2}).
We note that proofs of the GSL that are more general than the proof of
this paper have previously been given \cite{GSLproofs}; however, the
previous proofs used specific properties of black-hole spacetimes,
unlike our analysis. 

Finally, we note that, as discussed further at the end of Sec.\ 
\ref{s:discussion}, our results
can be generalized straightforwardly to arbitrary
spacetime dimensions greater than 2.   

\subsection{Derivations of entropy bound and of generalized second
law: framework, viewpoint and assumptions}

The starting point for our derivation of the entropy bounds
(\ref{Bousso1}) and (\ref{Bousso2}) is a postulated phenomenological
description of 
entropy, which differs from assumptions that have been used in the
past to derive the GSL \cite{GSLproofs}.  In this
section we describe our phenomenological description of entropy and
its motivation.

First, note that one of the hypotheses of Bousso's conjecture is the
dominant energy condition, which is often violated by the expected
stress energy tensor of matter in semiclassical gravity.  Hence the
conjecture cannot have the status of a fundamental law as it is
currently stated, but rather can
only be relevant in ``classical regimes'' where the dominant energy
condition is satisfied \cite{Lowe}.  It may be possible to replace the
dominant energy condition by a quantum inequality of the type invented
by Ford and Roman \cite{FR95,FRnew,flan97,FE98,FR99} to overcome this
difficulty\footnote{Lowe \cite{Lowe} argues that the Bousso conjecture
must fail for  
a system consisting of an evaporating black hole accreting at just the
right rate to balance the Hawking radiation mass loss.  For such a
system, it would seem that the black hole can accrete an arbitrary
amount of entropy without changing its area, and in addition it is
hard to see how a modified Bousso conjecture incorporating a quantum
inequality rather than a local energy condition could be satisfied.
However, this counterexample might be resolved by the fact that it
may be appropriate to assign a negative entropy flux at the horizon
to states with an outgoing Hawking flux, or it might be resolved by
making adjustments to the Bousso conjecture.}.  In this paper we will
assume the null convergence condition, that $T_{ab} 
k^a k^b \geq 0$ for all null vectors $k^a$ [see Eqs.\ (\ref{b1}) and 
(\ref{e1})--(\ref{e2}) below], which is weaker than the dominant
energy condition.   Thus, our proof of
Bousso's conjecture, like the conjecture itself, is limited to
``classical regimes'' in which local energy conditions are satisfied.

Clearly, in order to derive the bounds (\ref{Bousso1}) and
(\ref{Bousso2}), we must make some 
assumptions about entropy. The entropy that the conjecture refers to
presumably should include gravitational contributions. It seems
plausible that any gravitational
entropy flux through the null hypersurface $L$ will be associated with
a shearing of that hypersurface, which has 
the same qualitative effect in the Raychaudhuri equation [see
Eq.\ (\ref{raych}) below] as a matter stress-energy flux. 
Thus, it may
be possible to treat gravitational contributions to entropy in a
manner similar to the matter contributions. However, our present
understanding of quantum gravity is not sufficient to attempt to
meaningfully quantify the gravitational contributions to entropy.
Consequently, in our analysis below, we shall consider only the matter
contributions to entropy.

With regard to the matter contribution to entropy, for both the GSL
and the Bousso bound, there is an apparent tension between the fact
that these statements are supposed to have the status of fundamental
laws and the fact that entropy is a quantity whose definition is
coarse-graining dependent.  
However, this tension is
resolved by noting that the number of degrees of freedom should be an
upper bound for the entropy $S$, irrespective of choice of
coarse-graining \cite{Bou99c}. Equivalently, we may restrict attention
to the case where the matter is locally in thermal equilibrium (i.e.,
maximum entropy density for its given energy density); if the bound
holds in this case, it must hold in all cases.

We shall proceed by assuming that a phenomenological description of
matter entropy can be given in terms of an entropy flux 4-vector $s^a$. We
shall then postulate some properties of $s^a$. In fact, we shall
postulate two independent sets of hypotheses on $s^a$, each of which
will be sufficient to prove the bound (\ref{Bousso1}); the first set
of hypotheses also will suffice to prove the bound (\ref{Bousso2}).
Note that it is not a central goal of this paper to 
justify our hypotheses, although we do discuss some motivations
below.  Instead we shall merely observe that they appear to hold in
certain regimes.  Note also that, at a fundamental level, entropy is a
non-local quantity and so can be well described by a entropy flux
4-vector only in certain regimes and over certain scales.  This fact
is reflected in our hypotheses below.

The first of our two sets of hypotheses is very much in the spirit of
the original Bekenstein bound (\ref{S/E}). Suppose that one has a null
hypersurface, 
$L$, the generators of which terminate at a finite value
$\lambda_\infty$ of affine parameter $\lambda$.
Suppose that one puts matter in a box and drops it
through $L$ in such a way that the back end of the box crosses $L$ at
$\lambda_\infty$. Then, if a bound of the nature of Eq.\ (\ref{S/E}) holds,
the amount of entropy crossing $L$ should be limited by the energy
within the box and the box ``size''. The box size, in turn, would be
related to the affine parameter at which the front end of the box
crossed $L$. On the other hand, suppose that matter flowing through
$L$ near $\lambda_\infty$ were not confined by a box. Then there would
be no ``box size restriction'' on the entropy flux near
$\lambda_\infty$.  However, in order to have a larger entropy flux
than one could achieve when using a box, it clearly would be necessary to
put the matter in a state where the ``modes'' carrying the entropy
``spill over'' beyond $\lambda_\infty$. In that case, it is far from
clear that the entropy carried by these modes should be credited as
arriving prior to $\lambda_\infty$, so that they would count in the
entropy flux through $L$. In other words, it seems reasonable to
postulate that the entropy flux through $L$ cannot be higher than the
case where the matter is placed in a box whose back end crosses $L$ at
$\lambda_\infty$, and to consider a bound on this entropy flux of the
general form of Eq.\ (\ref{S/E}).

The above considerations motivate the following hypothesis concerning
the entropy flux.  We assume that associated with every null surface
$L$ there is an entropy flux 4-vector $s^a_L$ from which one can
compute the entropy flux through $L$. Let $\gamma$ be a null geodesic
generator of $L$, with affine parameter $\lambda$ and tangent $k^a =
(d / d\lambda)^a$.  If $\gamma$ is of infinite affine parameter
length, then $T_{ab} k^a k^b=0$ along $\gamma$ by the focusing theorem
\cite{Wald}, and we assume that $s^a_L=0$ along $\gamma$.
On the other hand, if $\gamma$ ends at a finite value, $\lambda_\infty$, of
affine parameter, then we assume that\footnote{From Eq.\
(\ref{raych}), our proof also works if we replace the 
hypothesis (\ref{b1}) with the 
weaker hypotheses that
$$
|s^a_L k_a | \le (\lambda_\infty - \lambda) \left[ \pi T_{ab} k^a k^b
+ {\hat \sigma}_{ab} {\hat \sigma}^{ab}/8 \right],
$$
where ${\hat \sigma}_{ab}$ is the shear tensor [Eq.\ (9.2.28) of Ref.\
\cite{Wald}] associated with the generators of $L$.
In this context, we can interpret $s^a_L$ to
be the combined matter and gravitational entropy flux, rather than
just the matter entropy flux.}
\beq
|s^a_L k_a | \le \pi (\lambda_\infty - \lambda) T_{ab} k^a k^b.
\label{b1}
\endeq 
The inequality (\ref{b1}) is a direct analog of the original
Bekenstein bound (\ref{S/E}), with
$|s^a_L k_a |$ 
playing the role of $S$, $T_{ab} k^a k^b$ playing the role of $E$, and
$\lambda_\infty - \lambda$ playing the role of $R$. As discussed
above, the motivation for the bound (\ref{b1}) is essentially the
same as that for the bound (\ref{S/E}). Note that Eq.\ (\ref{b1}) is
independent of the choice of 
affine parameterization of $\gamma$; i.e., both sides of this equation
scale the same way under a change of affine parameter.

The above set of hypotheses has the property that the entropy flux, $- s^a_L
k_a$, depends upon $L$ in the sense (described above) that modes that
only partially pass through $L$ prior to $\lambda_\infty$ do not
contribute to the entropy flux.  In our second set of hypotheses, we
assume the existence of an absolute entropy flux 4-vector $s^a$, which
is independent of the choice of $L$.
We assume that this $s^a$ obeys the following purely local, pointwise
inequalities for any null vector $k^a$: 
\beq
(s_a k^a)^2 \le \alpha_1 \, T_{ab} k^a k^b 
\label{e1}
\endeq
and
\beq
|k^a k^b \nabla_a s_b| \le \alpha_2 \, T_{ab} k^a k^b,
\label{e2}
\endeq
where $T_{ab}$ is the stress-energy tensor
\footnote{The stress energy
tensor appearing in these inequalities should be 
interpreted as a macroscopic or averaged stress energy tensor ${\bar
T}_{ab}$, rather than a microscopic stress energy tensor $T_{ab}$.
For example, for an atomic gas, the fundamental microscopic
stress-energy tensor $T_{ab}$ will vary rapidly over atomic and
nuclear scales, while a suitable averaged macroscopic stress tensor
${\bar T}_{ab}$ can be taken to vary only over macroscopic scales
(like the conventional entropy current $s^a$).  Thus our results apply
to null surfaces $L$ of an averaged, macroscopic metric ${\bar
g}_{ab}$ rather than the physical metric $g_{ab}$ \cite{Zal97}.  
Note that null surfaces of ${\bar g}_{ab}$ can differ significantly
from the null surfaces of $g_{ab}$, since with suitable microscopic
sources (for example cosmic strings) a null surface of $g_{ab}$ can be
made to intersect itself 
very frequently without the occurrence of caustics.  
However, the boundary of the future (or past) of the 2-surface $B$ with
respect to ${\bar g}_{ab}$ should be close to the boundary of the future
(respectively, past) of $B$ with respect to $g_{ab}$. Thus, if one wishes
to work with the exact metric $g_{ab}$, one should presumably replace the
null hypersurface, $L$, in the Bousso conjecture and our generalization
(\ref{Bousso2}) with a suitable portion of the boundary of the future (or
past) of $B$. This new formulation of the conjectures should hold whenever
Eq.\ (\ref{b1}) or Eqs.\ (\ref{e1}) and (\ref{e2}) hold for the
macroscopically averaged entropy current and macroscopically averaged
stress energy tensor.  An alternative
interpretative framework would be to assume the existence of an
``entropy current'' which varies rapidly on the smallest scales that are
compatible with our gradient assumption (\ref{e2}) (atomic and nuclear
scales in our example), in which case our result would apply directly
to the microscopic metric.}.  Here $\alpha_1$ and $\alpha_2$ can be any
positive constants that satisfy  
\beq
(\pi \alpha_1)^{1/4} + (\alpha_2 / \pi)^{1/2} =1.
\label{alphavals}
\endeq
[Recall that we are using Planck units with $G=c=\hbar=k=1$.]  A
specific simple choice of $\alpha_1$ and $\alpha_2$ that satisfy
the condition (\ref{alphavals}) is $\alpha_1 = 1/(16 \pi)$ and
$\alpha_2 = \pi/4$, which are the values quoted in the abstract above.
Note that, like Eq.\ (\ref{b1}), Eqs.\ (\ref{e1}) and (\ref{e2}) are
independent of the choice of scaling  of $k^a$. 
Also note that both of our sets of hypotheses (\ref{b1}) and
(\ref{e1})--(\ref{e2}) imply the null convergence condition $T_{ab}
k^a k^b \ge 0$, as mentioned above.

We now turn to a discussion of the physical regimes in which we expect
the pointwise assumptions (\ref{e1}) and (\ref{e2}) of our second set
of hypotheses to be valid.
The first assumption (\ref{e1}) of our second set of
hypotheses says, roughly speaking, that  
the entropy density is bounded above by the square root of the energy
density.  One can check 
that the condition is satisfied for thermal equilibrium states of Bose
and Fermi gases except at temperatures above a critical temperature of
order the Planck temperature \footnote{
Specifically, for a free massless boson gas at temperature $T$ the
stress energy tensor 
has the form $T_{ab} = (\rho+p) u_a u_b + p g_{ab}$ and the entropy
flux vector is $s^a = \sigma u^a$, where $p=\rho/3$ and $\sigma = 4
\rho / (3 T)$.  It follows that for any null vector $k^a$ we have
$(s_a k^a)^2 / T_{ab}k^ak^b = 4 \rho / (3 T^2) = 2 \pi^2 g N_s T^2 /
45$, where $g$ is the number of polarization components and $N_s$ is
the number of species.}.  One can also check that for quantum fields
in a box at low 
temperatures (the example discussed in Sec.\ \ref{s:background} above),
the condition (\ref{e1}) is violated only if the box is Planck size or
smaller, or if the number of species is allowed to be
very large.  Thus, it seems plausible that the
bound (\ref{e1}) will be universally valid if one assumes a Planck
scale cutoff for physics and if one also assumes a limit to the number
of species.  Also one can argue as follows that a bound of the form of
Eq.\ (\ref{e1}) should follow from the 
Bekenstein bound (\ref{S/E}).  Consider a
region of space of that is sufficiently small that (i) the entropy density
and energy density are approximately uniform over the region, and (ii)
the region is weakly self-gravitating so that its total energy $E$
satisfies $E \alt R$, where $R$ is the size of the region.  Then, if
$S$ is the total entropy in the region, the ratio of entropy density
squared to energy density is $\sim S^2 / E R^3 \le 4 \pi^2 E / R$ by Eq.\
(\ref{S/E}), which is $\alt 1$ as $E \alt R$.

The second assumption (\ref{e2}) states roughly that the 
gradient of the entropy density is bounded above by the energy
density. For a 
free, massless boson or fermion gas in local thermal equilibrium, this
condition reduces to the condition that the temperature gradient,
$|\nabla T|$, be small compared with $T^2$, i.e., that the fractional
change in $T$ over a distance $1/T$ be smaller than unity. This
condition must be satisfied in order for the notion of local thermal
equilibrium to make sense.

In addition, it would appear that condition (\ref{e2}) is necessary
for our entire phenomenological description of entropy as represented
by an 4-current $s^a$ to be valid.  To see this, consider the following
illustrative example.  Consider a wavepacket mode of a quantum
field, where the wavelength is $\lambda$ and where the volume occupied
by the wavepacket is $ f \lambda^3$ for some dimensionless factor $f
\agt 1$. 
Consider a state where this wavepacket mode is occupied by $N$
particles.  Such a system has a 
well defined expected stress energy tensor $\langle {\hat T}_{ab}
\rangle$, whose corresponding energy density will be of order 
\beq
\rho \sim {N \over f \, \lambda^4}.
\label{energydensity}
\endeq
We now imagine that we are to somehow model such
a system with a smooth entropy flux vector $s^a$.  We expect that
the total entropy carried by the system should be of order $N$, so
that the entropy density $s$ should be approximately
\beq
s \sim { N \over f\, \lambda^3}.
\label{entropydensity}
\endeq
Clearly the
concept of local entropy flux here cannot make sense on scales short
compared to the wavelength $\lambda$; only in an averaged sense, on
scales comparable to $\lambda$ or larger, does the concept of entropy
flux make sense.  Thus, the lengthscale ${\cal L} = s / | \nabla s|$
over which the entropy density varies should be greater than or of the
order of $\lambda$.  From the estimates (\ref{energydensity}) and
(\ref{entropydensity}), the condition ${\cal L} \agt \lambda$ is
equivalent to $|\nabla s| \alt \rho$, which is essentially our 
assumption (\ref{e2}).  Hence, our second condition
(\ref{e2}) rules out the class of entropy currents $s^a$ which vary
significantly over scales shorter than $\lambda$, allowing 
only the more appropriate $s^a$
that vary over scales of a wavelength or longer.

In summary, we expect our second set of hypotheses to be valid in regimes
where the following conditions are satisfied:  (i) Spacetime structure can
be accurately described by a classical metric, $g_{ab}$, and the
gravitational contributions to entropy, other than that from black holes,
are negligible.  (ii) The matter entropy can be accurately be described by
an entropy current $s^a$. In particular, this condition will be valid in
familiar hydrodynamic regimes. (iii) The vacuum energy contributions to
the stress-energy tensor are negligible, so that the stress-energy tensor
satisfies classical energy conditions.

We shall refer to regimes satisfying the above three conditions as
``classical'', even though, in such regimes, quantum physics may play an
essential role in accounting for the entropy of matter.  In classical
regimes, our hypotheses (\ref{e1}) and (\ref{e2}) should be valid.  We
have argued above that hypothesis (\ref{b1}) also should hold.  Hence our
arguments show that the Bousso bound (\ref{Bousso1}) and its
generalization (\ref{Bousso2}) should hold in classical regimes.
While our arguments do not 
show that the entropy bounds (\ref{Bousso1}) and (\ref{Bousso2}) hold
at any fundamental 
level, they do show that any counterexample either must involve quantum
phenomena in an essential way (in the sense of failure to be in a
classical regime), or must violate Eq.\ (\ref{b1}) and/or Eq.\
(\ref{e1}) or Eq.\ (\ref{e2}).

\section{DERIVATION OF ENTROPY BOUNDS}
\label{s:proof}

In this section we derive the generalized entropy bound
(\ref{Bousso2}) from the assumption (\ref{b1}) 
and the Bousso bound (\ref{Bousso1}) from the assumptions
(\ref{e1})--(\ref{e2}). 

We start with some definitions and constructions.  First, we can
without loss of generality take the vector field $k^a$ on the
2-surface $B$ to be future directed, since the conjecture is time
reversal invariant.   Let $l^a$ be the unique vector field on $B$
which is null, future directed, orthogonal to $B$ and which satisfies
$l^a k_a =-1$.  We extend both $k^a$ and $l^a$ to $L$ by parallel
transport along the null geodesic generators of 
$L$.  Thus, $k^a$ is tangent to each geodesic.  Then the expansion
$\theta = \nabla_a k^a$ is well defined on $L$ and independent of how
$k^a$ is extended off of $L$, since
\beq
\theta = (g^{ab} + k^a l^b + k^b l^a) \nabla_a k_b.
\endeq
By the hypotheses of Bousso's conjecture and of its generalization
(\ref{Bousso2}), $\theta$ is nonpositive
everywhere on $L$.
Let $\{x^\Gamma\} = (x^1, x^2)$ be any coordinate system on $B$.
Then one obtains a natural coordinate system $(\lambda, x^1, x^2)$ on
$L$ in the obvious way, where $k^a = (d / d \lambda)^a$ and we take
$\lambda =0$ on $B$.  

For the generator $\gamma$ which starts at the point $x^\Gamma$ on $B$, let 
$\lambda_\infty(x^\Gamma)$ be the value of affine parameter
(possibly $\lambda_\infty = \infty$) at the endpoint of the generator.
This endpoint can either be a caustic ($\theta = -\infty$) or have a
finite expansion $\theta$.
We can without loss of generality exclude the case $\lambda_\infty =
\infty$, since otherwise we must have $T_{ab} k^a k^b=0$ and
$\theta=0$ along
$\gamma$ by the focusing theorem, and then either version of our
hypotheses implies that $s^a =0$ along $\gamma$, so that there is 
no contribution to the entropy flux.  Thus, generators of infinite
affine parameter length make no contribution to the LHS of the
inequalities (\ref{Bousso1}) and (\ref{Bousso2}) while making a
non-negative contribution to 
the RHS, and so can be ignored.
For the generators of finite affine parameter length, 
we can without loss of generality rescale the affine parameter
along each generator in order to make the endpoint occur at
$\lambda_\infty=1$.

\subsection{Reducing the conjecture to each null geodesic generator}

Next, we show that it is sufficient to focus attention on each individual
generator of $L$, one at a time.  More specifically we have the
following lemma.

\medskip
\noindent
{\bf Lemma:}  {\it A sufficient condition for the generalized entropy
bound (\ref{Bousso2}) is 
that for each null geodesic generator $\gamma$ of $L$ of finite affine
parameter length, we have
\beq
\int_0^{1} \ d \lambda \ (- s_a k^a) {\cal A}(\lambda)
\ \le {1 \over 4} \left[1 - {\cal A}(1) \right],
\label{gcondt}
\endeq
where
\beq
{\cal A}(\lambda) \equiv \exp \left[ \int_0^\lambda d {\bar \lambda} \,
\theta({\bar \lambda}) \right]
\label{Adef}
\endeq
is an area-decrease factor associated with the given generator.
Similarly, a sufficient condition for the Bousso bound (\ref{Bousso1})
is that
\beq
\int_0^{1} \ d \lambda \ (- s_a k^a) {\cal A}(\lambda)
\ \le {1 \over 4}
\label{gcondt1}
\endeq
along each generator.}

\medskip

To prove the lemma, first note that by assumption the entropy flux
through $L$ is given by
\beq
S_L = \int_L s^a \epsilon_{abcd},
\label{si0}
\endeq
where the orientation on the hypersurface $L$ is that determined by
the 3-form 
\beq
{\hat \epsilon}_{bcd} \equiv l^a \epsilon_{abcd}.
\label{hatedef}
\endeq
Here we are using the notation of Appendix B of Ref.\ \cite{Wald}
for integrals of differential forms, and
we also use the abstract index notation of Ref.\ \cite{Wald}
for all Roman indices throughout the paper.
The formula (\ref{si0}) applies for either version of our
phenomenological description of entropy; i.e., we can use either $s^a$
or $s^a_L$ in the integrand.  In Appendix \ref{s:integralformula}
we derive the following formula for the integral (\ref{si0}) in the
coordinate system $(\lambda, x^\Gamma)$:
\beq
S_L = \int_B d^2x \ 
\sqrt{ {\rm det} \, h_{\Gamma\Lambda}(x)} \, 
\int_0^{\lambda_\infty(x)} \ d \lambda \ s(\lambda) {\cal A}(\lambda).
\label{intformula}
\endeq
Here $x \equiv (x^1,x^2) = x^\Gamma$, $h_{\Gamma\Lambda}(x)$ is the
induced 2-metric on the 
2-surface $B$, $\lambda_\infty(x)$ is the value of affine
parameter at the endpoint of the generator which starts at $x$,
and
\beq
s \equiv - s_a k^a.
\label{jdef}
\endeq
Note that $s$ is non-negative for future directed, timelike or null
$s^a$, which we expect to be the case.  (However, our proof does not
require the assumptions that $s^a$ be timelike or null and future
directed.)  Now as discussed above, $s(\lambda) =0$ for those generators
of infinite affine parameter length, so it follows that
\beq
S_L = \int^\prime_B d^2x \ 
\sqrt{ {\rm det} \, h_{\Gamma\Lambda}(x)} \, 
\int_0^{1} \ d \lambda \ s(\lambda) {\cal A}(\lambda),
\label{intformula1}
\endeq
where $\int^\prime_B d^2 x$ denotes an integral only over those
generators of finite affine parameter length.
Now we see that if the condition
(\ref{gcondt}) is satisfied, then we 
obtain from Eq.\ (\ref{intformula1}) that
\beq
S_L \le {1 \over 4} \int^\prime_B d^2x 
\sqrt{ {\rm det} \, h_{\Gamma\Lambda}(x)}  \left[1 -
{\cal A}(1,x) \right].
\label{intformula2}
\endeq
The generalized entropy bound (\ref{Bousso2}) now follows from Eq.\
(\ref{intformula2}), using the fact that 
the area $A_B$ of the 2-surface $B$ is given by
\beq
A_B = \int_B d^2x 
\sqrt{ {\rm det} \, h_{\Gamma\Lambda}(x)},
\label{ABformula}
\endeq
while the area $A_{B^\prime}$ of the 2-surface $B^\prime$ composed of
the endpoints of the generators is
\beq
A_{B^\prime} = \int^\prime_B d^2x 
\sqrt{ {\rm det} \, h_{\Gamma\Lambda}(x)} {\cal A}(1,x).
\label{ABprimeformula}
\endeq
Similar arguments show that the Bousso bound (\ref{Bousso1}) follows
from the assumption (\ref{gcondt1}).

\subsection{Preliminaries}

{}From the lemma, its sufficient now to prove the condition
(\ref{gcondt}) or, respectively, the condition (\ref{gcondt1})
for each finite-affine-parameter-length null generator $\gamma$ of $L$.  
For ease of notation we henceforth drop the dependence on the
$x^\Sigma$ coordinates in all quantities.  
Now the twist along each of the null generators will vanish, since it
is vanishing initially on the two surface $B$, and the evolution
equation for the twist \cite{Wald} then implies that it always vanishes.
The Raychaudhuri equation in the relevant case of vanishing twist can thus
be written as 
\beq
- { d \theta \over d \lambda} = {1 \over 2} \theta^2 + f(\lambda),
\label{raych}
\endeq
where $f = 8 \pi T_{ab} k^a k^b+ {\hat \sigma}_{ab} {\hat
\sigma}^{ab}$ and ${\hat \sigma}_{ab}$ is the shear tensor.  The
function 
$f$ is non-negative 
by the null convergence condition [which follows from either Eq.\
(\ref{b1}) or Eqs.\ (\ref{e1})--(\ref{e2})] since ${\hat
\sigma}_{ab} {\hat \sigma}^{ab} 
\ge 
0$ always.  The assumption (\ref{b1}) of our first set of hypotheses
now implies that 
\beq
|s(\lambda)|  \le (1 - \lambda) f(\lambda) /8.
\label{b1a}
\endeq
Similarly, our second set of hypotheses (\ref{e1}) and (\ref{e2})
implies that 
\beq
s(\lambda)^2 \le {\bar \alpha}_1 f(\lambda)
\label{e1a}
\endeq
and
\beq
| s'(\lambda) | \le {\bar \alpha}_2 f(\lambda),
\label{e2a}
\endeq
where 
\beq
{\bar \alpha}_1 = 8 \pi \alpha_1, \ \ \ \ {\bar \alpha}_2 = 8 \pi 
\alpha_2.
\label{baralphas}
\endeq
We define the quantity 
\beq
I_\gamma \equiv \int_0^1 d \lambda \, s(\lambda) {\cal A}(\lambda).
\label{Igammadef}
\endeq
Our tasks now are to show that $I_\gamma \le [1-{\cal A}(1)]/4$ when
Eq.\ (\ref{b1a}) holds, and that $I_\gamma \le 1/4$ when Eqs.\
(\ref{e1a}) and (\ref{e2a}) hold, using
only the definition (\ref{Adef}) of the 
area-decrease factor and the Raychaudhuri equation (\ref{raych}).

Now by assumption any geodesic generator must terminate no later than
the point (if it exists) at which ${\cal A}(\lambda) \to 0$.  Hence we
have ${\cal A}(\lambda) \ge 0$ 
everywhere on $L$.  
It is convenient to define $G(\lambda) = \sqrt{{\cal A}(\lambda)}$, from
which it follows from the definition (\ref{Adef}) of the area-decrease factor
and from the Raychaudhuri equation (\ref{raych}) that
\beq
f(\lambda) = - 2 {G^{\prime\prime}(\lambda) \over G(\lambda)}.
\label{ff}
\endeq
It follows that $G^{\prime\prime}$ is negative.  Also the expansion
$\theta$ is always negative, and hence $G^\prime$ is always negative,
so that $G$ is monotonically decreasing, starting at the value
$G(0)=1$, and ending at some value $G(1)$ with $0 \le G(1) \le 1$.
In particular, we have $0 \le G(\lambda) \le 1$ for all $\lambda$.
For those generators which terminate at caustics we have $G(1) =
{\cal A}(1)=0$, but not all generators will terminate at caustics; some might
terminate at the auxiliary spacelike 2-surface $B^\prime$.

\subsection{Proof of the generalized Bousso bound under the first set
of hypotheses}

Using the formula (\ref{b1a}) 
and the definition $G = \sqrt{{\cal A}}$ we
find that the integral (\ref{Igammadef}) satisfies
\beq
I_\gamma \le {1 \over 8} \int_0^1 d \lambda \, (1 - \lambda)
f(\lambda) G(\lambda)^2.
\endeq
{}From the formula (\ref{ff}) for $f(\lambda)$, this can be written as
\beq
I_\gamma \le - \int_0^1 d \lambda \, (1 - \lambda)
G^{\prime\prime}(\lambda) G(\lambda) / 4.
\label{ii}
\endeq
Now we have $0 \le G(\lambda) \le 1$, so we can drop the factor of
$G(\lambda)$ in the integrand of Eq.\ (\ref{ii}).  Integrating by
parts and using the fundamental theorem of calculus now gives 
\beq
I_\gamma \le {1 \over 4} \left[ G(0) - G(1) + G^\prime(0)\right].
\label{240a}
\endeq
Now 
\beq
G(0) - G(1) = 1 - G(1) \le 1 - {\cal A}(1),
\endeq
since $G(0)=1$ and $G(1) = \sqrt{{\cal A}(1)} \ge {\cal A}(1)$.
Also the third term in Eq.\ (\ref{240a}) is negative.  It follows that
\beq
I_\gamma \le {1 \over 4} \left[ 1 - {\cal A}(1) \right],
\endeq
as required.

\subsection{Proof of the original Bousso bound under the second set of
hypotheses}

First, note that without loss of generality we can assume that
the function $s(\lambda)$ is nonnegative.   This is because we can replace
$s(\lambda)$ by $|s(\lambda)|$ in the integral (\ref{Igammadef})
without decreasing the value of the integral, and
the assumptions (\ref{e1a}) and (\ref{e2a}) are satisfied by the
function $|s|$ if they are satisfied by $s$, since $| \, |s|^\prime\,|
\le |s^\prime|$.

We start by fixing a $\lambda_1$ in $(0,1)$, the value of which we
will pick later.  
We then choose a $\lambda_0$ in $[0,\lambda_1]$ which minimizes $f$ in
the interval $[0,\lambda_1]$; i.e., we choose a $\lambda_0$ which satisfies
\beq
f(\lambda_0) = \min_{0 \le \lambda \le \lambda_1} f(\lambda).
\endeq
[We assume that the function $f(\lambda)$ is continuous so that this 
minimum is attained].\footnote{The proof extends easily to the
non-continuous case.  If we choose
$\epsilon>0$ and choose $\lambda_0$ so that 
$$
f(\lambda_0) = (1 + \epsilon) \, {\rm g.l.b.}\ \left\{ f(\lambda)
\right| 0 \le \lambda \le \lambda_1 \} ,
$$
then we can come as close as we please to satisfying the inequality
(\ref{finalans}).  Hence the inequality (\ref{finalans}) is
satisfied.}  We now show that 
\beq
f(\lambda_0) \le {\pi^2 \over 2 \lambda_1^2} \left[ 1 - {\cal A}(1) \right].
\label{fbound}
\endeq
To see this, let $\theta_0(\lambda)$ and ${\cal A}_0(\lambda)$ be the
expansion and area-decrease factor that would be 
obtained by solving the Raychaudhuri equation (\ref{raych}) with
$f(\lambda)$ 
replaced by $f(\lambda_0)$ and using the same initial condition 
$\theta_0(0) = \theta(0)$.  Since $f(\lambda) \ge f(\lambda_0)$ for
$0 \le \lambda \le \lambda_1$, it is clear that we must have
\beq
{\cal A}(\lambda) \le {\cal A}_0(\lambda)
\label{al}
\endeq
for $0 \le \lambda \le \lambda_1$.  But the explicit solution of the
Raychaudhuri equation for $\theta_0(\lambda)$ and ${\cal A}_0(\lambda)$ is
\beq
\theta_0(\lambda) = - \sqrt{2 f(\lambda_0)} \tan \left[ \sqrt
{ {f(\lambda_0) \over 2} }(\lambda + {\hat \lambda}) \right]
\endeq
and
\beq
{\cal A}_0(\lambda) = { \cos^2 \left[ \sqrt{f(\lambda_0) \over 2} (\lambda +
{\hat \lambda}) \right] \over
\cos^2 \left[ \sqrt{f(\lambda_0) \over 2}
{\hat \lambda} \right]},
\endeq
where ${\hat \lambda}$ is a constant in $[0,1]$.
Applying the inequality (\ref{al}) at $\lambda = \lambda_1$ now yields
\beq
{\cal A}(\lambda_1) \le {\cal A}_0(\lambda_1) \le \cos^2 \left[ \sqrt
{ f(\lambda_0) \over 2} \lambda_1 \right].
\label{intf}
\endeq
Using the inequality $\sin \chi \ge 2 \chi / \pi$ which
is valid for $0 \le \chi \le \pi/2$, and the inequality ${\cal A}(1) \le
{\cal A}(\lambda_1)$, one can obtain the upper bound (\ref{fbound}) from Eq.\
(\ref{intf}).

Next, we split the integral (\ref{Igammadef}) into a contribution
$I_1$ from the interval $[0,\lambda_0]$ and a contribution $I_2$ from
the interval $[\lambda_0,1]$:
\beq
I_\gamma = \int_0^{\lambda_0} \, s {\cal A} + \int_{\lambda_0}^1 \, s
{\cal A} = I_1
+ I_2.
\endeq
In the formula for $I_1$, we drop the factor of ${\cal A}$ which is $\le 1$,
insert a factor of $1 = d\lambda / d\lambda$, and integrate by
parts to obtain  
\beq
I_1 \le I_{1b} + I_1^\prime.
\endeq
Here $I_{1b}$ is the boundary term that is generated, given by
\beq
I_{1b} = s(\lambda_0) \lambda_0,
\label{I1bdef}
\endeq
and 
\beq
I_1^\prime =  - \int_0^{\lambda_0} d\lambda \,
s^\prime(\lambda) \lambda.
\label{I1primedef}
\endeq
Similarly we insert a factor of $1 = d(\lambda-1) / d\lambda$ into the
formula for $I_2$ and integrate by parts, which yields
\beq
I_2 = I_{2b} + I_2^\prime,
\endeq
where the boundary term is
\beq
I_{2b}  = s(\lambda_0) {\cal A}(\lambda_0) (1 - \lambda_0)
\label{I2bdef}
\endeq
and where
\beq
I_2^\prime = \int_{\lambda_0}^1 d \lambda \left[ s^\prime {\cal A} (1 -
\lambda) + s {\cal A}^\prime (1 - \lambda) \right].
\label{I2primedef}
\endeq
An upper bound on the total integral is now given by the relation
\beq
I_\gamma \le I_{1b} + I_{2b} + I_1^\prime + I_2^\prime.
\label{split}
\endeq
We now proceed to derive bounds on the integrals $I_1^\prime$,
$I_2^\prime$ and on the total boundary term $I_{1b} + I_{2b}$.

Consider first the total boundary term, which from Eqs.\ (\ref{I1bdef}) and
(\ref{I2bdef}) is given by
\beq
I_{1b} + I_{2b} = s(\lambda_0) \left[ \lambda_0 + (1 - \lambda_0)
{\cal A}(\lambda_0) \right].
\endeq
Since ${\cal A}(\lambda_0)$ and $\lambda_0$ both lie in $[0,1]$, this is
bounded above by $s(\lambda_0)$.  If we now use our assumption
(\ref{e1a}), we find
\beq
I_{1a} + I_{1b} \le \sqrt{{\bar \alpha}_1 \, f(\lambda_0)},
\endeq
and using the bound (\ref{fbound}) on $f(\lambda_0)$ finally yields
\beq
I_{1a} + I_{1b} \le 
{ \sqrt{{\bar \alpha}_1} \pi \over \sqrt{2}
\lambda_1} \left[ 1 - {\cal A}(1) \right]^{1/2}.
\label{boundaryans}
\endeq

Turn now to the integral $I_1^\prime$.  Inserting the assumption
(\ref{e2a}) into the formula (\ref{I1primedef}) for $I_1^\prime$ and
using the formula 
(\ref{ff}) for $f(\lambda)$ yields
\beq
I_1^\prime \le - 2 {\bar \alpha}_2 \int_0^{\lambda_0} { G^{\prime
\prime} \over G} \lambda \, d \lambda.
\endeq
Since $G$ is a decreasing function, we can replace the $1/G(\lambda)$
in the integrand by $1/G(\lambda_0)$.  If we then integrate by parts
and use the fundamental theorem of calculus, we obtain
\beq
I_1^\prime \le {2 {\bar \alpha}_2 \over G(\lambda_0) } \left
[ G(\lambda_0) - 1 - G^\prime(\lambda_0) \lambda_0 \right].
\endeq
Since $G(\lambda_0) \le 1$ this yields
\beq
I_1^\prime \le - {2 {\bar \alpha}_2 G^\prime(\lambda_0) \lambda_0
\over G(\lambda_0) }.
\label{I1primeans1}
\endeq

We now show that for all $\lambda$ in $[0,1]$,
\beq
- {G^\prime(\lambda) \over G(\lambda)} \le {1 \over 1 - \lambda}
\left[ 1 - {\cal A}(1) \right].
\label{lemma}
\endeq
To see this, apply the mean value theorem to the function $G$ over the
interval $[\lambda,1]$, which yields
\beq
G(1) - G(\lambda) = (1 - \lambda) G^\prime(\lambda_*),
\endeq
for some $\lambda_*$ in $[\lambda,1]$.  But since $G^{\prime\prime}$
is negative by Eq.\ (\ref{ff}), we have $G^\prime(\lambda_*) \le
G^\prime(\lambda)$, and it follows that
\begin{eqnarray}
- {G^\prime(\lambda) \over G(\lambda)} &\le& 
{1 \over 1 - \lambda}
\left[ 1 - {G(1) \over G(\lambda)} \right] \nonumber \\
\mbox{} 
&\le& 
{1 \over 1 - \lambda} \left[ 1 - G(1) \right] 
\le {1 \over 1 - \lambda} \left[ 1 - {\cal A}(1) \right].
\end{eqnarray}
Here the last inequality follows from $G = \sqrt{{\cal A}}$ and $0 \le
G \le 1$.  Using
the relation (\ref{lemma}), our upper bound (\ref{I1primeans1}) for 
$I_1^\prime$ now yields
\beq
{I_{1}^\prime \over 1 - {\cal A}(1)} \le 2 {\bar \alpha}_2 {\lambda_0 \over 1 - \lambda_0} \le 2 {\bar \alpha}_2 {\lambda_1 \over 1 -\lambda_1}.
\label{I1primeans}
\endeq

Finally, we turn to the integral $I_2^\prime$.  The second term in the
formula (\ref{I2primedef}) for $I_1^\prime$ is negative and so can be
dropped.  In the 
first term, we use the formula $G = \sqrt{{\cal A}}$, the formula (\ref{ff})
for $f(\lambda)$ and our gradient assumption in the form (\ref{e2a})
to obtain 
\beq
I_2^\prime \le - 2 {\bar \alpha}_2 \int_{\lambda_0}^1 d \lambda \,
G^{\prime\prime} G (1 - \lambda).
\endeq
Now since $0 \le G(\lambda) \le 1$ for all $\lambda$, we can drop the
factor of $G(\lambda)$ in the integrand.
If we then integrate by parts and use the fundamental theorem of
calculus we obtain 
\beq
I_{2}^\prime \le 2 {\bar \alpha}_2 \left[ G(\lambda_0) - G(1) + (1 - \lambda_0)
G^\prime(\lambda_0) \right].
\label{240}
\endeq
Now since $G(\lambda_0) \le 1$ and $G = \sqrt{{\cal A}}$ we have
\beq
G(\lambda_0) - G(1) \le 1 - G(1) \le 1 - {\cal A}(1).
\endeq
Also, the last
term in Eq.\ (\ref{240}) is negative.  Hence we obtain the upper bound
\beq
I_{2}^\prime \le 2 {\bar \alpha}_2 \left[ 1 - {\cal A}(1) \right].
\label{I2primeans}
\endeq

Finally we combine Eq.\ (\ref{split}) with 
the upper bounds (\ref{boundaryans}),
(\ref{I1primeans}), and (\ref{I2primeans}) 
for the boundary term $I_{1b} + I_{2b}$ and for the integrals
$I_1^\prime$ and $I_{2}^\prime$ to yield
\begin{eqnarray}
I_\gamma &\le& \sqrt{{\bar \alpha}_1} { \pi \over \sqrt{2}
\lambda_1} \left[ 1 - {\cal A}(1) \right]^{1/2} 
+ 2 {\bar \alpha}_2 {1 \over 1 - \lambda_1}
\left[ 1 - {\cal A}(1) \right] \nonumber \\
\mbox{} &\le& 
\left[ \sqrt{{\bar \alpha}_1} { \pi \over \sqrt{2} \lambda_1} 
+ 2 {\bar \alpha}_2 {1 \over 1 - \lambda_1} \right] \ 
\left[ 1 - {\cal A}(1) \right]^{1/2}.
\end{eqnarray}
Choosing the value of $\lambda_1$ that minimizes this upper bound
yields
\beq
I_\gamma \le \left[\sqrt{ \pi \sqrt{{\bar \alpha}_1 \over 2}} +
\sqrt{2 {\bar \alpha}_2}\right]^2 \ \left[ 1 - {\cal A}(1)
\right]^{1/2}. 
\endeq
Using the definition (\ref{baralphas}) of the 
the parameters ${\bar \alpha}_1$ and ${\bar \alpha}_2$ together with
the assumption (\ref{alphavals}) yields
\begin{eqnarray}
I_\gamma &\le& { 1 \over 4} \left[ 1 - {\cal A}(1)\right]^{1/2}
\nonumber \\
\mbox{} &\le& {1 \over 4},
\label{finalans}
\end{eqnarray}
as required.
Note that our proof actually implies the inequality
\beq
S_L \le {1 \over 4} A_B^{1/2} \left[ A_B - A_{B^\prime}
\right]^{1/2}.
\endeq
This inequality is stronger than the Bousso bound
(\ref{Bousso1}) but weaker than the generalized Bousso bound (\ref{Bousso2}).

\section{CONCLUSION}
\label{s:discussion}

We have shown that the generalization (\ref{Bousso2}) of 
Bousso's entropy bound is satisfied under 
the hypothesis (\ref{b1}), and that the original Bousso bound
(\ref{Bousso1}) holds under the hypotheses (\ref{e1}) -- (\ref{e2}).
While these hypotheses are unlikely to represent 
relations in any fundamental theory, they appear to be satisfied for 
matter in a certain semi-classical regime below the Planck scale.  As
such, our results rule out a large class of possible counterexamples to 
Bousso's conjecture, including cases involving gravitational collapse
or other strong gravitational interactions.  As with Bousso's bound, 
if the holographic principle is indeed part of a fundamental theory, it
may be that the hypotheses discussed here will provide clues to 
its formulation.

Note that we do not show in this paper that the entropy bounds
(\ref{Bousso1}) and (\ref{Bousso2}) can be saturated by entropy
4-currents $s^a$ 
satisfying our assumptions.  However, consideration of simple
examples shows that the bound (\ref{Bousso1}) comes within a factor of order
unity of being saturated by currents satisfying our second set of
hypotheses (\ref{e1}) and (\ref{e2}).
Also, a simple scaling argument shows that the least upper bound
on the ratio $S_L/A_B$ for currents satisfying our
assumptions is of the form $\alpha_2 F(\alpha_1 \alpha_2^{-2})$
for some function $F$.  As a result, at fixed $\alpha_1 \alpha_2^{-2}$
the least upper bound depends continuously on $\alpha_2$.  This
guarantees that there exist some values of $\alpha_1$ and $\alpha_2$,
of order unity, such the entropy bound (\ref{Bousso1}) is both 
satisfied and can be saturated by entropy currents satisfying the
inequalities (\ref{e1}) and (\ref{e2}).  A similar statement is true
for the bound (\ref{Bousso2}) and the hypothesis (\ref{b1}).  
 
In our analysis above, we have taken the dimension of spacetime to
be 4. However, in an $n$-dimensional spacetime with $n > 2$, the
Raychaudhuri equation continues to take the form (\ref{raych}), except that the
coefficient of $\theta^2$ on the right side is now
$1/(n-2)$. Consequently, if we define $G = {\cal A}^{1/(n-2)}$ in the
$n$-dimensional case, an equation of the form (\ref{ff}) will continue to hold
with the factor of $2$ on the right side replaced by $(n-2)$.
The remainder of our analysis can then be carried out in direct parallel
with the 4-dimensional case. Thus, with suitable adjustments to the
numerical factors appearing in Eqs.\ (\ref{b1}), (\ref{e1}), and
(\ref{e2}), all of our results continue to hold for all spacetime
dimensions greater than 2.

\acknowledgements

We thank Warren Anderson and Raphael Bousso for helpful conversations,
and the Institute for Theoretical Physics (ITP) in Santa Barbara,
where this work was initiated, for its hospitality.
We also thank Jacob Bekenstein for discussions on the validity of
the entropy bound (\ref{S/E}) at low temperatures.
We particularly wish to thank Raf Guedens for bringing to our
attention an error in an earlier version of this manuscript.
EF acknowledges the support of the Alfred P. Sloan Foundation.  This
work has been supported by NSF grants PHY 9407194 to the ITP, 
PHY 95-14726 to the University of Chicago,
PHY 9722189 to Cornell University, 
PHY 97-22362 to to Syracuse University,
and by funds provided by Syracuse University.

\appendix
\section{formula for integral over null hypersurface}
\label{s:integralformula}

In this appendix we derive the formula (\ref{intformula}) for the
integral 
\beq
S_L = \int_L s^a \epsilon_{abcd}
\label{si}
\endeq
of the entropy current over the null hypersurface $L$ [cf.\ Eq.\
(\ref{si0}) above], using the coordinate system $(\lambda, x^\Gamma) =
(\lambda, x^1,x^2)$ defined in Sec.\ \ref{s:proof}.

We start by discussing the relation between tensors on the spacetime $M$ and
tensors on the null surface $L$.  We introduce the notation that
capital roman indices $A,B,C, \ldots$ denote tensors
on $L$, in the sense of the abstract index convention of Ref.\
\cite{Wald}.  For any 1-form $w_a$ defined on $L$, we will denote the   
pullback of $w_a$ to $L$ as 
\beq
w_A = P_A^a w_a;
\endeq
this defines the operator $P_A^a$.  
Since $k_a$ is normal to $L$, the pullback of $k_a$ vanishes, so $P^a_A
k_a = 0$.
Using the null tetrad introduced at the beginning of Sec.\ II, we can
define a similar mapping between vectors on $M$ and vectors on $L$.
At any point ${\cal P}$ in $L$, the projection
operation 
\beq
v^a \to (\delta^a_b + l^a k_b) v^b
\label{projection1}
\endeq
maps the 4-dimensional tangent space $T_{\cal P}(M)$ into the
3-dimensional tangent space $T_{\cal P}(L)$.  Thus one can write the
mapping (\ref{projection1}) as $v^a \to v^A = Q^A_a v^a$, which
defines the operator $Q^A_a$.  
Note that the vector $l^a$ is annihilated by the projection operation
(\ref{projection1}), while $k^a$ and vectors perpendicular to $k^a$
and $l^a$ are unchanged.  We define $k^A = Q^A_a k^a = (d /
d\lambda)^A$, which is the tangent vector in $L$ the generators of $L$.

Consider now the integrand in the integral (\ref{si}).  It is
proportional to 
\beq
s^a \, k_{[a} l_b e_c f_{d]},
\label{3f}
\endeq
where $e^a$ and $f^a$ are spacelike vector fields such that
$\{ k^a, l^a, e^a, f^a \}$ is an orthonormal basis.
When we pullback the 3-form (\ref{3f}) to $L$, all the terms where
the index on $k_a$ is free and not contracted with $s^a$ will be
annihilated.  Hence without loss of generality we can replace $s^a$
with $- (s_b k^b) l^a$.  Thus 
we obtain from
Eqs.\ (\ref{si}) and (\ref{hatedef}) that
\beq
S_L = \int_L s \, {\hat \epsilon}_{abc},
\label{SL1}
\endeq
where $s = - s_a k^a$ and where the 3-form ${\hat \epsilon}_{abc}$ is
defined in Eq.\ (\ref{hatedef}) above.

As a tool for evaluating the integral (\ref{SL1}), we define an
induced connection $D_A$ on $L$ by 
\beq
D_A v^B = P^c_A \, Q^B_d \, \nabla_c v^d
\label{conn}
\endeq
where $v^d$ is any vector field on $M$ with $v^A = Q^A_b v^b$.  
One can check that this formula defines a derivative operator on $L$.
Next we note that the
pullback ${\hat \epsilon}_{ABC}$ of the 3-form (\ref{hatedef}) is parallel
transported along each null generator of $L$ with respect to the
connection (\ref{conn}):
\beq
k^A D_A \, {\hat \epsilon}_{BCD} =0.
\label{pt}
\endeq
This follows from the fact that $k^a$, $l^a$ and $\epsilon_{abcd}$ are
parallel transported along each generator with respect to the
4-dimensional connection\footnote{Note however that unlike the
situation in the four dimensional setting, in general 
$
D_A \ {\hat \epsilon}_{BCD} \ne 0; 
$
i.e., ${\hat \epsilon}_{ABC}$ is not covariantly constant
with respect to $D_A$.}.  Next, consider the Lie derivative ${\cal
L}_{\vec k} {\hat \epsilon}$ 
of ${\hat \epsilon}_{ABC}$ with respect to $k^A$.  Since the result is
a 3-form we must have 
\beq
\left({\cal L}_{\vec k} {\hat \epsilon}\right)_{ABC} = \eta \, {\hat
\epsilon}_{ABC},
\label{ansatz}
\endeq
for some scalar field $\eta$.  We can define a upper index volume form
${\hat \epsilon}^{ABC}$ by the requirement that
\beq
{\hat \epsilon}^{ABC} {\hat \epsilon}_{ABC} = 3!.
\endeq
[A definition is terms of raising indices is inapplicable here since
there is no natural non-degenerate metric on $L$.]
Now contracting both sides of Eq.\ (\ref{ansatz}) with ${\hat
\epsilon}^{ABC}$ and using Eq.\ (\ref{pt}) yields
\beq
\eta = D_A k^A = \nabla_a k^a,
\endeq
which is just the usual expansion $\theta$.

Next, we define a 3-form ${\tilde \epsilon}_{ABC}$ on $L$ by
demanding that it coincide with ${\hat \epsilon}_{ABC}$ on the
2-surface $B$, and that it be Lie transported along the generators of
$L$.  If we write ${\hat \epsilon}_{ABC} = \zeta {\tilde \epsilon}_{ABC}$,
where $\zeta$ is a scalar field on $L$, it follows from
Eq. (\ref{ansatz}) with $\eta = \theta$ that 
\beq
{\cal L}_{\vec k} \zeta = \theta \zeta.
\endeq
Solving this equation using the definition (\ref{Adef}) of the
area-decrease factor yields $\zeta = {\cal A}$.  Thus we see that the
geometrical  
meaning of the factor ${\cal A}$ is that it is the ratio between the
Lie-transported 3-volume form ${\tilde \epsilon}_{ABC}$ and the
parallel transported 3-volume form ${\hat \epsilon}_{ABC}$, where in
both cases one starts from the 2-surface $B$.

Consider now the specific coordinate system $(\lambda, x^\Gamma) =
(\lambda, x^1, x^2)$.  In this coordinate system the fact that
${\tilde \epsilon}_{ABC}$ is Lie transported along the generators
translates into
\beq
{\partial \over \partial \lambda} {\tilde \epsilon}_{\lambda x^1 x^2}(\lambda,x) =0,
\endeq
so that 
\beq
{\tilde \epsilon}_{\lambda x^1 x^2}(\lambda,x) = 
{\tilde \epsilon}_{\lambda x^1 x^2}(0,x) = 
\sqrt{ {\rm det} \, h_{\Gamma\Lambda}(x) },
\endeq
where 
${\tilde \epsilon}_{\lambda x^1 x^2}(\lambda,x)$ denotes one of
the coordinate components of the tensor ${\tilde \epsilon}_{ABC}$ in
the coordinate system $(\lambda,x^\Gamma)$, $x \equiv (x^1,x^2)$ as
before, and $h_{\Gamma\Lambda}$ is the induced 2-metric on $B$.  It
follows that  
\beq
{\hat \epsilon}_{\lambda x^1 x^2}(\lambda,x) = {\cal A}(\lambda,
x) \, \sqrt{ {\rm det} \, h_{\Gamma\Lambda}(x) }.
\label{3fans}
\endeq
Combining this with the formula (\ref{SL1}) for the entropy flux
finally yields the formula (\ref{intformula}).


\end{document}